\documentclass[letter]{aa}

\usepackage{graphicx}
\usepackage{txfonts}
\usepackage{booktabs}

\begin{document}

\title{Sound propagation in loosely confined rubble-pile surfaces}

\author{Eric S. Frizzell\inst{1}\fnmsep\thanks{Corresponding author: ericscott.frizzell@polimi.it}
\and Irina L. San Sebasti\'an\inst{1}
\and Fabio Ferrari\inst{1}}

\institute{Department of Aerospace Science and Technology, Politecnico di Milano, 20156, Milano, Italy}

\date{Received September 30, 20XX}

\abstract
{Rubble-pile asteroids are weakly confined granular aggregates whose seismic response may differ from expectations based on bulk elastic properties. Near a free surface, confining pressures may fall to the order of Pa or less.}
{We examine the acoustic-to-nonlinear transition in very weakly compressed granular material and assess its implications for mechanical disturbances at rubble-pile asteroid surfaces.}
{We combine analytical wave-speed relations for precompressed Hertzian chains with discrete-element simulations, then map the one-dimensional relations to surface preloads estimated from gravitational overburden on spherical small bodies.}
{The chain recovers an approximately constant acoustic branch and a nonlinear branch close to the expected one-fifth-power scaling of wave speed with impact velocity. The transition impact velocity follows the predicted static-overlap dependence and falls much faster than the acoustic speed as confinement weakens, leaving both low acoustic speeds and a narrow acoustic regime.}
{The one-dimensional Hertzian framework indicates that weak confinement can collapse the acoustic regime itself, so low interaction velocity alone is insufficient to justify an acoustic description of rubble-pile surfaces.}

\keywords{regolith --
rubble-pile asteroids --
granular materials --
acoustic waves --
shock waves --
discrete element method}

\maketitle
\nolinenumbers

\section{Introduction}
\label{sec:introduction}

The mechanical response of rubble-pile asteroids remains difficult to predict because their internal structures and constituent materials are poorly constrained. These bodies are gravitational aggregates of blocks, grains, voids, and weak contacts, and their near-surface material may be highly porous, fractured, and only weakly confined by self-gravity \citep{walsh2018rubble,bagatin2020gravitational,raducan2020effects,molaro2020insitu,brisset2022asteroid,barnouin2024geology}. As a result, seismic velocities inferred or assumed for small bodies span orders of magnitude, from a few meters per second in weakly confined granular material and simulated rubble piles to kilometers per second for coherent-rock or megaregolith analogues \citep{garcia2015micro,murdoch2017probing,matsue2020measurements,nishiyama2021simulation,sanchez2022transmission,dellagiustina2024seismology,gallot2025seismic}.

Under weak confinement, however, wave propagation cannot always be described by a single amplitude-independent sound speed. In Hertzian granular media, weak disturbances propagate through a precompressed contact network with an acoustic speed controlled by the static overlap, whereas stronger disturbances enter an amplitude-dependent nonlinear regime approaching the sonic-vacuum limit \citep{nesterenko2001dynamics,coste1997solitary,coste1999validity,daraio2005strongly,wildenberg2013shock,gomez2012shocks}. Moreover, disordered three-dimensional packings introduce additional effects like force-chain branching, scattering, contact opening, and network rearrangement, which generally reduce and redistribute propagation relative to the ordered chain \citep{owens2011sound,clark2015nonlinear}. The crossover becomes especially important near free surfaces, where the confining pressure and static contact overlap rapidly decrease.

This regime is directly relevant to spacecraft interactions with small-body surfaces. Low-velocity contacts by Hayabusa, MASCOT, and OSIRIS-REx revealed highly mobile and weak near-surface material on Itokawa, Ryugu, and Bennu \citep{yano2006touchdown,scholten2019descent,brisset2020regolith,lauretta2022spacecraft,walsh2022nearzero}, motivating the question of whether apparently gentle surface interactions remain within the acoustic regime. \citet{murdoch2021lowvelocity} similarly argued that the velocity separating quasi-static and fluid-like granular interaction should decrease substantially under asteroid gravity. We consider the corresponding question for disturbance propagation through the contact network. The one-dimensional Hertzian chain provides a useful reference because it isolates these contact-controlled regimes and their scaling with weak confinement.

Here we use Hertzian particle-chain theory and discrete-element simulations to characterize the acoustic-to-nonlinear transition and apply this framework to weakly confined rubble-pile surfaces. We show that the transition velocity decreases much more rapidly with diminishing preload than the acoustic wave speed, causing the acoustic regime to collapse near a free surface.

\section{Methods}
\label{sec:methods}

\subsection{Hertzian acoustic and nonlinear wave speeds}
\label{sec:wave_speed_eqns}

We model particle-particle contacts using the nonlinear elastic contact law of \citet{hertz1882contact}. For the spherical-particle contacts used in our analytical model, the elastic normal force is
\begin{equation}
\mathbf{F}_{n}
=
\frac{4}{3}E^*\sqrt{r_p^*}\,\delta_n^{3/2}\mathbf{n},
\label{eq:normal_elastic_hertz}
\end{equation}
where $E$ is Young's modulus, $r_p$ is the particle's radius, $\delta_n$ is the normal overlap and $\mathbf{n}$ is the contact normal. For identical spheres, the equivalent properties become $r_p^*=r_p/2$ and $E^*=E/[2(1-\nu^2)]$, with $\nu$ being Poisson's ratio.

For a one-dimensional chain of identical Hertzian spheres, the long-wavelength sound speed from \citet{nesterenko2001dynamics} is
\begin{equation}
c_0=
\left(\frac{3\sqrt{2}}{\pi}\right)^{1/2}
\sqrt{\frac{E^*}{\rho_p}}
\left(\frac{\delta_0}{r_p}\right)^{1/4},
\label{eq:1Dsoundspeed}
\end{equation}
where $\rho_p$ is the particle density and $\delta_0$ is the quasi-static contact overlap. Thus, the acoustic speed decreases with decreasing precompression and approaches zero in the sonic-vacuum limit.

When precompression becomes negligible, propagation instead depends on disturbance amplitude. For a Hertzian chain in the sonic-vacuum limit, a particle in the leading wavefront traveling at non-linear speed $c_N$ experiences maximum force $F_{max}$. From \citet{nesterenko2001dynamics} and \citet{coste1997solitary}, $c_N\propto F_{\max}^{1/6}$ and $F_{\max}\propto v_0^{6/5}$, which we use to write
\begin{equation}
c_N=
\xi\sqrt{\frac{E^*}{\rho_p}}
\left(
\frac{v_0}{\sqrt{E^*/\rho_p}}
\right)^{1/5},
\label{eq:1Dshocks}
\end{equation}
where $\xi$ is a dimensionless prefactor that sets the normalization of the nonlinear branch and is determined from the particle-chain response. Impact velocity, $v_0$, is the corresponding disturbance.

Finite precompression introduces a crossover between these limits, written here as the transition velocity, $v_t$: the imposed impact velocity $v_0$ at which the acoustic (Eq.~\ref{eq:1Dsoundspeed}) and nonlinear wave-speed (Eq.~\ref{eq:1Dshocks}) relations intersect.
\begin{equation}
v_{\mathrm{t}}=
\xi^{-5}\sqrt{\frac{E^*}{\rho_p}}
\left(\frac{3\sqrt{2}}{\pi}\right)^{5/2}
\left(\frac{\delta_0}{r_p}\right)^{5/4}.
\label{eq:1Dtransition}
\end{equation}

Equivalently, $v_t\propto c_0(\delta_0/r_p)$, such that the transition occurs when the dynamic strain becomes comparable to the static contact strain \citep{sen2008solitary}. Note that $v_t$ is a crossover scale, not a sharp boundary \citep{jiao2023revisiting}. In the following analysis, we denote the fitted acoustic- and nonlinear-branch exponents by $q_{\rm ac}$ and $q_{\rm NL}$, respectively, and the fitted transition-scaling exponent by $\alpha_t$. The scaling $v_t\propto(\delta_0/r_p)^{5/4}$ is especially important under weak confinement: as $\delta_0$ decreases, the acoustic speed falls as $(\delta_0/r_p)^{1/4}$, while the transition velocity decreases more rapidly.

\subsection{Particle-chain model}
\label{sec:chain_model}

We simulate the particle chain using the discrete element method (DEM, \citet{cundall1979discrete}) implemented with the smooth-contact Hertz model in Project Chrono \citep{tasora2016chrono}. The normal elastic contact response follows Eq.~\eqref{eq:normal_elastic_hertz}, with viscous damping included in the numerical contact model; we give the remaining implementation details in Appendix~\ref{app:chain_numerics}.

Following \citet{frizzell2023simulation}, we arrange 30 identical spherical particles in a one-dimensional chain with the two end particles fixed. We prescribe the static overlap $\delta_0$ by setting the center-to-center spacing to $2r_p-\delta_0$, allow the chain to settle, and reuse the settled state for the impact simulations.

We initiate each simulation by assigning the first mobile particle an axial impact velocity $v_0$. We restrict $v_0$ to remain well below the particle-material sound speed and use it to set the conservative contact-resolution timestep described in Appendix~\ref{app:chain_numerics}.

We measure wave arrival independently from the particle-velocity magnitude $v_p$, the mean normal contact overlap $\delta$, and the particle pressure $p$. For $v_p$, we use the first qualifying post-impact local maximum; for $\delta$ and $p$, we use the dominant positive post-impact disturbance relative to the initial state. We then calculate the local propagation speed from the corresponding arrival times using the centered finite difference
\begin{equation}
c_i
=
\frac{y_{i+1}-y_{i-1}}
     {t_{i+1}-t_{i-1}},
\label{eq:chain_local_speed}
\end{equation}
where $y_i$ and $t_i$ are the axial position and arrival time of particle $i$. We evaluate all three diagnostics at $i=3$, using particles $i=2$ and $4$ (zero-based indexing). Appendix~\ref{app:chain_numerics} details the diagnostics and acceptance criteria.

\subsection{Transition characterization and surface extrapolation}
\label{sec:test_plan}

We construct a database of 200 precompressed one-dimensional Hertzian-chain configurations spanning the material and surface conditions in Table~\ref{tab:chain_test_matrix}. For each configuration, we sweep the impact velocity from the nonlinear regime toward lower amplitudes until we resolve the acoustic plateau and refine the sampling around the transition. We use the resulting wave-speed curves to determine $v_t$, test its predicted dependence on precompression, and constrain the nonlinear prefactor $\xi$. Appendix~\ref{app:transition_fitting} gives the sampling, fitting, and uncertainty-estimation procedures.

\begin{table}
\centering
\caption{Parameter ranges sampled in our Hertz-chain database.}
\label{tab:chain_test_matrix}
\begin{tabular}{lcc}
\hline\hline
Parameter & Symbol & Range \\
\hline
Young's modulus & $E$ & $5~\mathrm{MPa}$ -$100~\mathrm{GPa}$ \\

Particle radius & $r_p$ & $0.05$--$5~\mathrm{m}$ \\

Poisson ratio & $\nu$ & $0.20$--$0.40$ \\

Body radius & $R_b$ & $50$--$5000~\mathrm{m}$ \\

Bulk density & $\rho_b$ & $1000$--$2500$ kg/m$^{3}$ \\

Packing fraction & $\phi$ & $0.55$--$0.90$ \\
\hline
\end{tabular}
\end{table}

To map these configurations to weakly confined asteroid surfaces, we estimate the local gravitational overburden for a spherical body of radius $R_b$ and uniform bulk density $\rho_b$. Neglecting rotational unloading, the pressure at shallow depth $h$ is
\begin{equation}
p(h)
\simeq
\rho_b g_s h
=
(4\pi/3)G\rho_b^2R_bh ,
\label{eq:surface_overburden_pressure}
\end{equation}
where $g_s=(4\pi/3)G\rho_bR_b$. Neglecting rotation maximizes the confinement and therefore gives a conservative upper estimate of the near-surface preload.

We use the local overburden pressure from Eq.~\eqref{eq:surface_overburden_pressure} to define a representative axial preload $F_0=p\pi r_p^2$. For each surface condition, this preload is imposed uniformly along the chain; gravity is not included in the chain dynamics. Inverting the contact law gives the corresponding static overlap,
\begin{equation}
\delta_0
=
\left[
\frac{3F_0}
{4E^*\sqrt{r_p^*}}
\right]^{2/3}.
\label{eq:chain_surface_overlap}
\end{equation}
This pressure--force mapping is an idealization used to prescribe the one-dimensional preload; it does not include coordination-number or three-dimensional contact-network corrections. At the smallest predicted $\delta_0$, grain roughness, non-Hertzian behavior \citep{goddard1990nonlinear}, and cohesion \citep{scheeres2010scaling} may become important, so these estimates should be interpreted as idealized Hertzian reference limits.

For the nominal near-surface calculation, we set $h=r_p$ and derive $\rho_p=\rho_b/\phi$. The adopted stiffness range includes compliant DEM end members \citep{demartini2019using} through stiff rock-like particles; returned Bennu particles have measured elastic moduli of approximately $12$--$16~\mathrm{GPa}$ \citep{ryan2026thermal}. We provide the complete case inventory as machine-readable supplementary material. Equations~(\ref{eq:1Dsoundspeed})--(\ref{eq:1Dtransition}) along with Eq.~\eqref{eq:chain_surface_overlap} then give the Hertz-chain surface estimates used below.

\section{Results}
\label{sec:results}

\subsection{Acoustic-to-nonlinear transition in the particle chain}
\label{sec:chain_transition}

\begin{figure}[t]
\centering
\includegraphics[width=\columnwidth]{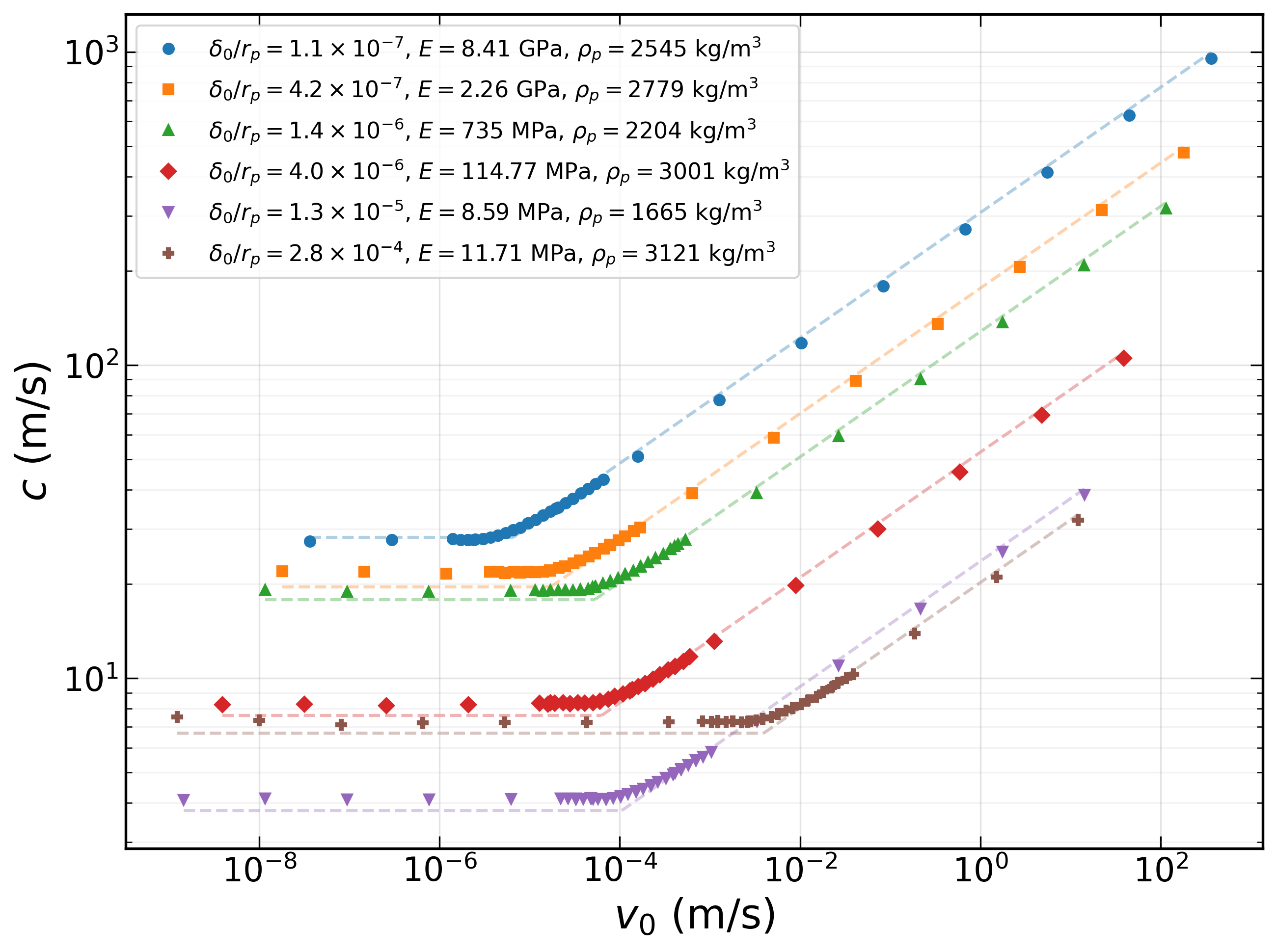}
\caption{Representative Hertzian-chain wave speeds measured via $v_p$ versus imposed impact velocity; each point is the result of an individual numerical simulation. Each configuration approaches the precompression-controlled acoustic speed at low $v_0$ and transitions toward amplitude-dependent nonlinear propagation at larger $v_0$. Dashed curves show the analytical acoustic and nonlinear branches from Eqs.~\eqref{eq:1Dsoundspeed} and \eqref{eq:1Dshocks}, joined at $v_t$ from Eq.~\eqref{eq:1Dtransition} using the diagnostic-specific nonlinear prefactor $\xi$.}
\label{fig:chain_transition}
\end{figure}

Across 200 particle-chain configurations, the simulations confirm the crossover predicted by Eqs.~(\ref{eq:1Dsoundspeed})--(\ref{eq:1Dtransition}) from acoustic propagation at low impact velocities to the Hertzian $c\propto v_0^{1/5}$ scaling at high velocities. Figure~\ref{fig:chain_transition} shows representative cases with the analytical branches overlaid. Each measurement gives the initial speed obtained from Eq.~\eqref{eq:chain_local_speed} using particle-velocity tracking.

Fits to the $v_p$, $\delta$, and $p$ diagnostics confirm this limiting behavior. All three give $q_{\rm NL}\simeq0.194$, close to the Hertzian prediction of $1/5$. The particle-velocity and pressure diagnostics also give approximately flat acoustic branches, $q_{\rm ac}\simeq0$. The fitted values and bootstrap intervals are reported in Appendix~\ref{app:transition_fitting}.

\subsection{Nonlinear normalization and transition scaling}
\label{sec:chain_collapse}

We determine the nonlinear prefactor $\xi$ in Eq.~\eqref{eq:1Dshocks} independently for each diagnostic. Table~\ref{tab:xi_metric_fits} reports the jointly fitted $\xi$ and $q_{\rm NL}$ values and the median $\xi$ obtained with $q_{\rm NL}=1/5$.

\begin{table}
\centering
\caption{Nonlinear-wave prefactor inferred from the three wave-arrival diagnostics. The free-slope value of $\xi$ is obtained together with $q_{\rm NL}$ from the global nonlinear-branch fit; brackets give the 95\% case-resampled bootstrap confidence interval. The final column gives the median case value when the Hertzian exponent $q_{\rm NL}=1/5$ is imposed.}
\label{tab:xi_metric_fits}
\begin{tabular}{lcc}
\hline\hline
Diagnostic & $\xi$ (free $q_{\rm NL}$) & Median $\xi$ ($q_{\rm NL}=1/5$) \\
\hline
Velocity & $0.911$ $[0.903,0.919]$ & $0.974$ \\
Overlap & $0.874$ $[0.864,0.883]$ & $0.935$ \\
Pressure & $0.837$ $[0.821,0.850]$ & $0.899$ \\
\hline
\end{tabular}
\end{table}

The three diagnostics yield similar nonlinear exponents but different normalizations, possibly reflecting their distinct peak-arrival definitions and near-source pulse development. Since $v_t\propto\xi^{-5}$, these differences shift the crossover velocity without changing its precompression dependence. We test that dependence by fitting the normalized transition velocity as a function of $\delta_0/r_p$. Appendix Fig.~\ref{fig:chain_collapse} shows that all three diagnostics recover $\alpha_t$ consistent with the Hertzian prediction of $5/4$ within the 95\% case-resampled bootstrap confidence intervals.

\subsection{Implications for weakly confined asteroid surfaces}
\label{sec:surface_bounds}

For the nominal surface extrapolation, we use the particle-velocity value of $\xi$ obtained with $q_{\rm NL}=1/5$. This is consistent with Eq.~\eqref{eq:1Dtransition}, while the free-$q_{\rm NL}$ fit provides an independent validation of the Hertzian exponent. The overlap- and pressure-derived values provide a diagnostic sensitivity range, with the pressure diagnostic giving the largest $v_t$.

\begin{figure}[t]
\centering
\includegraphics[width=0.95\columnwidth]{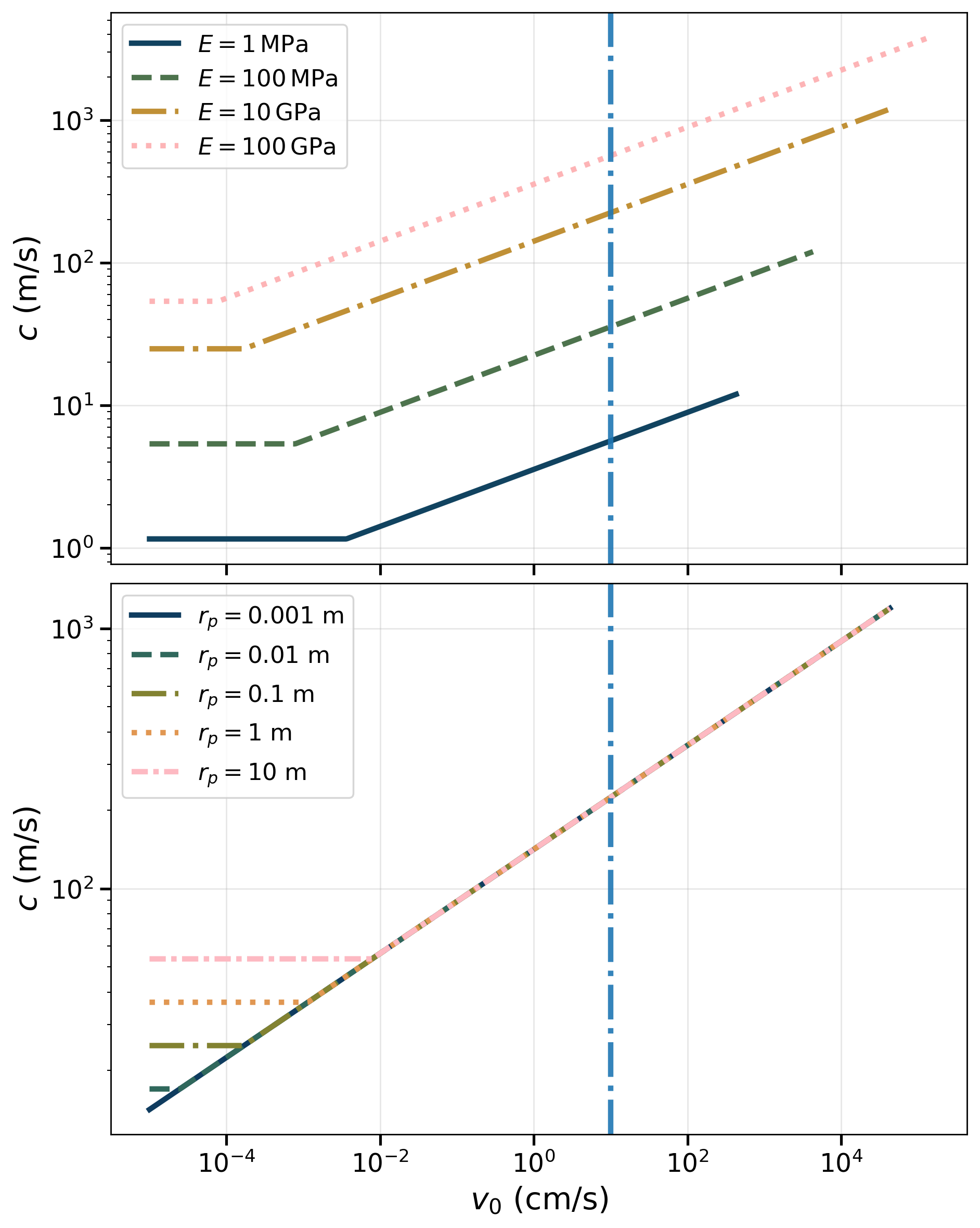}
\caption{Analytical Hertz-chain wave-speed estimates under weak surface confinement for varying $E$ (top) and $r_p$ (bottom). We use nominal Bennu-like properties ($E$ = 10 GPa, $r_p$ = 10 cm) unless varied in the legend. The vertical dash-dotted line marks a representative spacecraft--surface interaction velocity
of 10 cm/s.}
\label{fig:surface_parameter_sweeps}
\end{figure}

Figure~\ref{fig:surface_parameter_sweeps} shows how the analytical surface wave-speed estimates vary with particle stiffness and size. Increasing $E$ raises the acoustic speed $c_0$ but lowers $v_t$, widening the separation between the acoustic and nonlinear regimes. Particle size enters through the adopted near-surface depth $h=r_p$: larger particles sample greater overburden and therefore larger static preload. Under this approximation, $c_0\propto r_p^{1/6}$ and $v_t\propto r_p^{5/6}$.

\begin{figure}[t]
\centering
\includegraphics[width=0.95\columnwidth]{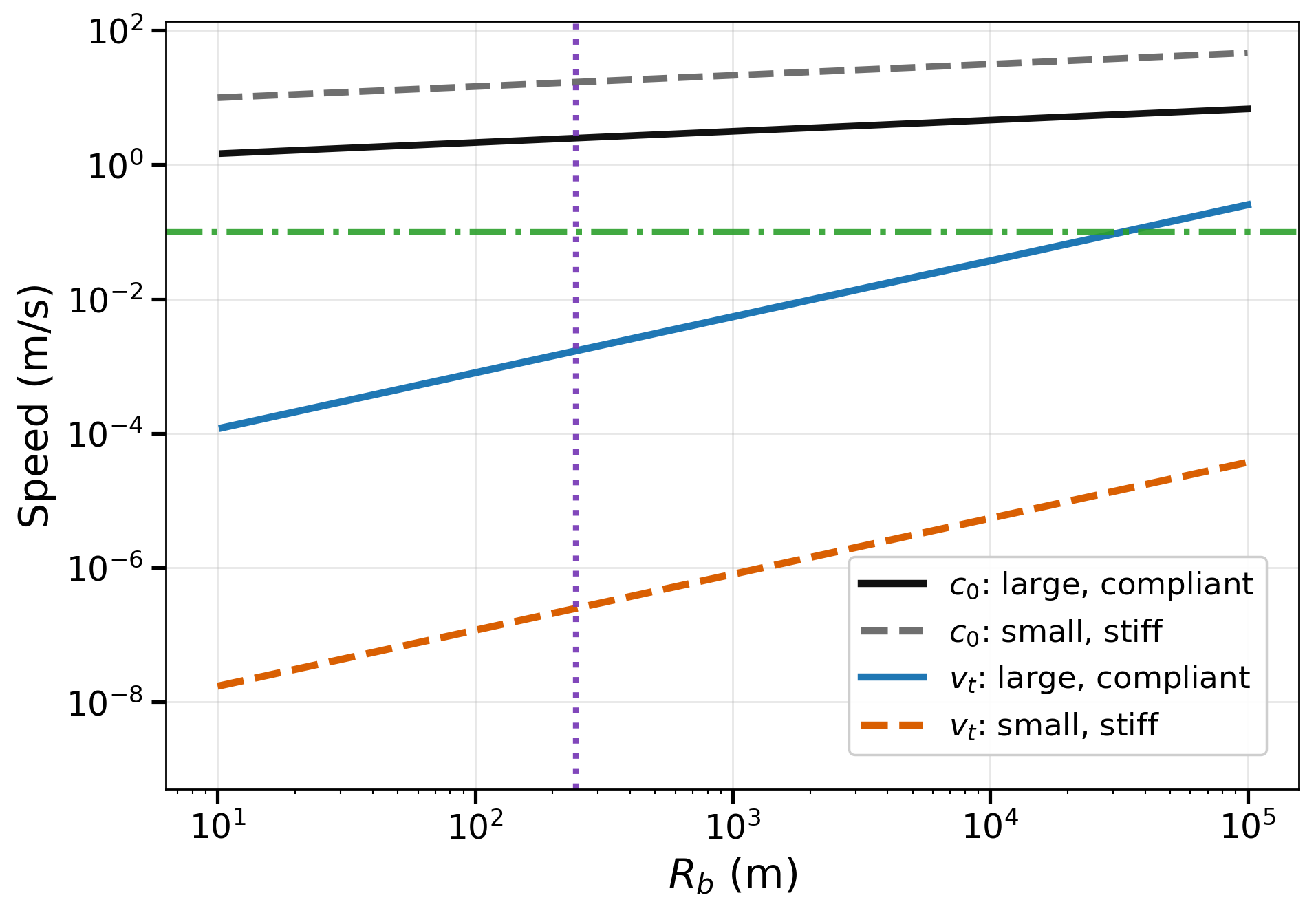}
\caption{Predicted near-surface acoustic speed $c_0$ and transition velocity $v_t$ as functions of body radius for compliant, meter-scale ($E=1$ MPa, $r_p=10$ m) and stiff, centimeter-scale ($E=10$ GPa, $r_p=1$ cm) end members. Note that these quantities are distinct; $c_0$ is a propagation speed while $v_t$ is the imposed impact velocity. Separate line styles and colors identify $c_0$ and the particle-velocity-diagnostic $v_t$ for the two cases, as indicated in the legend. The vertical line marks Bennu, and the horizontal line indicates an interaction velocity of 10 cm/s. The overlap- and pressure-derived transition curves are similar and are omitted for clarity; the pressure diagnostic yields the most conservative $v_t$.}
\label{fig:surface_bounds}
\end{figure}

Figure~\ref{fig:surface_bounds} shows the corresponding dependence on body size for compliant, meter-scale and stiff, centimeter-scale end members. Across the sampled range, the predicted near-surface acoustic speed remains below a few tens of meters per second, while the transition velocity is substantially smaller. Near Bennu-like conditions, even the compliant, meter-scale end member gives $v_t$ roughly two orders of magnitude below a representative spacecraft--surface interaction velocity of $10$ cm/s; for the stiff, centimeter-scale end member, the separation exceeds five orders of magnitude. Thus, a disturbance may remain far below the constituent-material sound speed while already lying well within the strongly nonlinear granular regime.

\section{Discussion and conclusions}
\label{sec:discussion}

The Hertz-chain results show that weak confinement reduces both the acoustic speed and the range of impact velocities over which propagation remains acoustic. Exceeding the transition impact velocity indicates nonlinear propagation through the contact network, but does not necessarily imply permanent grain rearrangement or surface modification, which also depend on packing, friction, impact geometry, and attenuation. In three-dimensional rubble piles, nonlinear propagation is expected to be most pronounced near the interaction site, where amplitudes are largest and confinement is weakest.

The one-dimensional equal-sphere relations provide a contact-scale reference rather than quantitative rubble-pile wave speeds. Experiments and simulations demonstrate \textit{slower} propagation in three-dimensional media than in ordered chains \citep{elshourbagy2008acoustics,cheng2013granular,manjunath2014plane,santibanez2016mechanical,frizzell2023simulation}. Quantitative application therefore requires independent three-dimensional calibration of both acoustic speed and transition impact velocity.

More generally, these results emphasize that the elastic modulus adopted in granular numerical models is not merely a numerical stiffness parameter when wave propagation or low-amplitude surface motion is of interest. Through the contact law, elastic modulus influences the quasi-static overlap and therefore both the acoustic wave speed and the transition into nonlinear propagation. Simulations intended to resolve seismic propagation or spacecraft--regolith interaction should therefore employ mechanically plausible contact properties, or demonstrate that the quantities of interest are insensitive to the adopted stiffness.

While previous three-dimensional DEM simulations have similarly identified impact-velocity-controlled propagation at low confinement \citep{sanchez2022transmission}, to our knowledge, this is the first application of the acoustic-to-nonlinear Hertzian transition framework to rubble-pile surface confinement.  Rather than treating weak confinement only as a reduction in sound speed, we show that it also restricts the range over which an acoustic description applies. As confinement decreases, this regime collapses toward very small disturbance velocities, so nominally gentle interactions may already propagate nonlinearly through the granular contact network. Low interaction velocity alone is therefore insufficient to justify an acoustic description of weakly confined rubble-pile surfaces, with implications for how seismic propagation and surface response are interpreted on regolith-covered small-body surfaces.

\begin{acknowledgements}

We acknowledge EU Horizon Europe funding under grants 101264707 (MSCA SEISMOR; EF, FF) and 101077758 (ERC TRACES; EF, IS, FF). Views expressed are solely those of the authors and do not necessarily reflect those of the EU or granting authorities, which bear no responsibility for them. We thank the anonymous reviewer for their helpful comments.


\end{acknowledgements}

\bibliographystyle{bibtex/aa}
\bibliography{references}

\begin{appendix}
\nolinenumbers 

\section{Particle-chain numerical implementation}
\label{app:chain_numerics}

We implement the particle chain in Project Chrono \citep{tasora2016chrono} using the smooth-contact Hertz model. The normal contact force is
\begin{equation}
\mathbf{F}_n
=
\left[
\frac{4}{3}E^*\sqrt{r_p^*}\,\delta_n^{3/2}
-
\gamma_n v_n
\right]\mathbf{n},
\label{eq:appendix_dem_normal_force}
\end{equation}
where $v_n$ is the normal relative velocity and $\gamma_n$ is the normal damping coefficient corresponding to the prescribed coefficient of restitution. We adopt a fixed restitution coefficient of $\epsilon=0.5$, following \citet{frizzell2023simulation}; sensitivity to damping is not examined here. Because the chain motion and loading are axial, tangential contact forces do not contribute.

We restrict the imposed impact velocity to $v_0\leq0.2c_m$, where \begin{equation}
c_m=\sqrt{\frac{E}{\rho_p}},
\end{equation} so that the imposed velocity remains well below the characteristic elastic-wave speed of the particle material. We set the integration timestep to $0.5\%$ of the elastic Hertzian collision time \citep{nesterenko2001dynamics,schwartz2012implementation},
\begin{equation}
\Delta t = 0.005t_c,\qquad \mathrm{where}\qquad
t_c =
2.94
\left[
\frac{(m^*)^2}
{r_p^*(E^*)^2v_{\rm rel}}
\right]^{1/5},
\label{eq:appendix_hertz_collision_time}
\end{equation}
where $m^*$ is the equivalent particle mass and $v_{\rm rel}$ is the relative collision velocity. We set $v_{\rm rel}=v_0$ and evolve each impact simulation for $60t_c$. The timestep is recalculated for each imposed impact velocity, including the smallest $v_0$, maintaining 200 integration steps per estimated collision time.

For the overlap diagnostic, we use the mean normal contact overlap associated with particle $i$,
\begin{equation}
\delta_i
=
\frac{1}{N_{c,i}}
\sum_j \delta_{n,ij},
\label{eq:appendix_mean_overlap}
\end{equation}
where $N_{c,i}$ is the number of contacts associated with that particle.

We compute the per-particle Cauchy stress tensor following
\citet{sanchez2012dem,zhang2017creep},
\begin{equation}
\boldsymbol{\sigma}_i
=
\frac{1}{V_i}
\sum_j
\mathbf{f}_{ij}\otimes\boldsymbol{\ell}_{ij},
\label{eq:appendix_particle_stress}
\end{equation}
where $V_i$ is the particle volume, $\mathbf{f}_{ij}$ is the contact force, and $\boldsymbol{\ell}_{ij}$ is the branch vector from the particle center to the contact point. We define the pressure diagnostic as
\begin{equation}
p_i
=
\frac{1}{3}
\left|
\operatorname{tr}\boldsymbol{\sigma}_i
\right|.
\label{eq:appendix_particle_pressure}
\end{equation}

For the particle-velocity diagnostic, we define arrival from the first qualifying local maximum in the post-impact velocity magnitude. For overlap and pressure, we identify the dominant positive post-impact disturbance relative to the initial state. 
Equation~\eqref{eq:chain_local_speed} is evaluated only where the neighboring particles have finite and sequentially increasing arrival times. Measurements that do not satisfy these requirements do not define a physically meaningful local propagation speed and are therefore left undefined.

\section{Transition fitting and uncertainty estimation}
\label{app:transition_fitting}

We sample the independent physical inputs with a deterministic low-discrepancy Halton sequence to obtain space-filling coverage of the multidimensional parameter range. We sample $E$, $r_p$, and $R_b$ logarithmically and $\nu$, $\rho_b$, and $\phi$ linearly. We derive $\rho_p=\rho_b/\phi$ and the static overlap from Eq.~\eqref{eq:chain_surface_overlap}. For each physical configuration, we sample $v_0$ logarithmically across the acoustic and strongly nonlinear regimes and refine the sampling around the provisional transition.

We characterize the transition independently from particle velocity, contact overlap, and pressure. We represent the low-amplitude branch by the analytical precompressed-chain sound speed, $c=c_0$, and the strongly nonlinear branch by
\begin{equation}
\frac{c}{\sqrt{E^*/\rho_p}}
=
\xi
\left(
\frac{v_0}{\sqrt{E^*/\rho_p}}
\right)^{1/5}.
\label{eq:appendix_nonlinear_fit}
\end{equation}
For each possible division of the ordered $c(v_0)$ measurements, we evaluate the logarithmic residual sum of squares of the two branches. The minimum defines the branch division, and their intersection gives $v_t$. We retain cases whose low-velocity branch differs substantially from the analytical acoustic prediction separately as suspect rather than including them in the nominal fits. These cases occur at the smallest static overlaps, where the acoustic plateau approaches the sonic-vacuum limit and becomes increasingly difficult to resolve numerically.

\begin{figure*}[t]
\centering
\includegraphics[width=\textwidth]{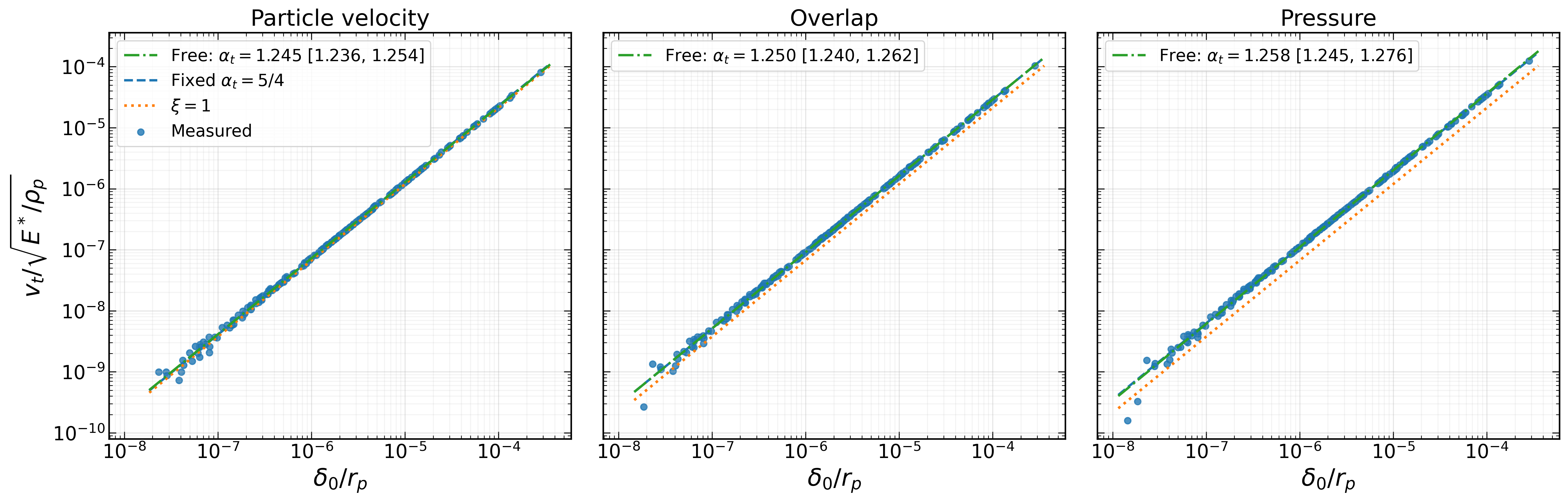}
\caption{Dimensionless acoustic-to-nonlinear transition velocity as a function of static precompression for $v_p$, $\delta$, and $p$. Points show the measured transitions; curves show free-$\alpha_t$ and fixed-$\alpha_t=5/4$ fits. The $\xi=1$ curve gives the no-fit Hertzian reference. Bracketed values in the labels give the 95\% case-resampled bootstrap confidence intervals on $\alpha_t$.}
\label{fig:chain_collapse}
\end{figure*}

We test the transition scaling with
\begin{equation}
\frac{v_t}{\sqrt{E^*/\rho_p}}
\propto
\left(
\frac{\delta_0}{r_p}
\right)^{\alpha_t},
\label{eq:appendix_transition_fit}
\end{equation}
where $\alpha_t$ is the fitted transition exponent. We evaluate both a free-$\alpha_t$ fit and a fit with the Hertzian prediction $\alpha_t=5/4$. Figure~\ref{fig:chain_collapse} compares the measured transition velocities with the free-$\alpha_t$, fixed-$\alpha_t=5/4$, and $\xi=1$ Hertzian reference relations.

We independently test the limiting branch slopes using the run-level data. For the acoustic branch, we fit
\begin{equation}
\frac{c}{c_0}
\propto
\left(
\frac{v_0}{v_t}
\right)^{q_{\rm ac}},
\end{equation}
with expected slope $q_{\rm ac}=0$. For the nonlinear branch, we fit
\begin{equation}
\frac{c}{\sqrt{E^*/\rho_p}}
=
\xi
\left(
\frac{v_0}{\sqrt{E^*/\rho_p}}
\right)^{q_{\rm NL}},
\end{equation} with both $\xi$ and $q_{\rm NL}$ free; the Hertzian prediction is $q_{\rm NL}=1/5$. The small deviation from the asymptotic Hertzian exponent may reflect finite-amplitude effects, viscous damping, and pulse development near the source. We estimate uncertainties in all fitted quantities using a case-resampled bootstrap. Each realization draws the same number of physical configurations as the original database, with replacement, while retaining all impact-velocity runs associated with each selected configuration. We repeat the complete fitting procedure for each realization and report the central 95\% interval of the resulting bootstrap distribution, defined by the 2.5th and 97.5th percentiles. These intervals quantify sensitivity to the sampled population of physical configurations rather than numerical uncertainty in an individual simulation. The fitted branch slopes and their uncertainties are summarized in Table~\ref{tab:vt_branch_fits}.

\begin{table}
\centering
\caption{Limiting-branch slopes obtained from the three wave diagnostics. Bracketed values give the 95\% case-resampled bootstrap confidence intervals. The Hertzian predictions are $q_{\rm ac}=0$ and $q_{\rm NL}=1/5$.}
\label{tab:vt_branch_fits}
\begin{tabular}{lcc}
\hline\hline
Diagnostic & $q_{\rm ac}$ & $q_{\rm NL}$ \\
\hline
Velocity & $0.001$ $[0.000,0.001]$ & $0.194$ $[0.193,0.195]$ \\
Overlap & $-0.024$ $[-0.029,-0.019]$ & $0.194$ $[0.193,0.195]$ \\
Pressure & $0.001$ $[-0.000,0.002]$ & $0.194$ $[0.192,0.195]$ \\
\hline
\end{tabular}
\end{table}

\end{appendix}

\end{document}